\documentclass[
aps,
prd,
reprint,
longbibliography, 
amsmath, 
amssymb, 
nofootinbib,
superscriptaddress
]{revtex4-2}

\usepackage[colorlinks=true, allcolors=black]{hyperref}
\usepackage{enumitem}
\usepackage[normalem]{ulem}
\usepackage{comment}
\usepackage[super]{nth}
\usepackage{graphicx}
\usepackage{dcolumn}
\usepackage{bm}
\usepackage[dvipsnames]{xcolor}
\usepackage{pgfplotstable, booktabs}
\usepackage{array}

\usepackage{amsmath}
\usepackage{xcolor}

\usepackage{float}
\usepackage{enumitem}

\newcommand\einstein{\texttt{Einstein Toolkit}}
\newcommand\ilgrmhd{\texttt{IllinoisGRMHD}}

\renewcommand{\d}[1]{\ensuremath{\operatorname{d}\!{#1}}}

\UseRawInputEncoding

\begin{document}

\title{Binary black holes gone MAD: Magnetically arrested minidisks around nonspinning black holes}

\author{Vikram Manikantan}
\email{vik@arizona.edu}
\affiliation{Steward Observatory \& Department of Astronomy, University of Arizona, Tucson, Arizona 85721, USA}

\author{Vasileios Paschalidis}
\email{vpaschal@arizona.edu}
\affiliation{Steward Observatory \& Department of Astronomy, University of Arizona, Tucson, Arizona 85721, USA}
\affiliation{Department of Physics, University of Arizona, Tucson, Arizona 85721, USA}
\date{\today}

\begin{abstract}
 We demonstrate the formation of magnetically arrested minidisks (MAM) around equal-mass, nonspinning binary black holes with magnetohydrodynamic simulations of circumbinary disk accretion in full 3+1 general relativity. The initial separation of $d\sim 30\,M$ allows the black holes to host large minidisks that suppress the total rest-mass accretion rate variability, which is modulated primarily at $\sim 1.6 \, f_{\rm orb}$. Each black hole horizon saturates with dimensionless magnetic flux $\phi \sim 30$. Magnetic reconnection near the horizons drives recurrent eruptions which are expected to drive flaring in the infrared and X-ray bands. Our results establish MAMs as a new outcome of circumbinary disk accretion, and a promising source of novel electromagnetic counterparts to gravitational waves from binary black holes.
\end{abstract}

\maketitle

\section{Introduction}\label{sec:intro}
Supermassive binary black holes (SMBBHs) are expected to form ubiquitously in the aftermath of galaxy mergers, and often in gaseous environments~\cite{begelman_massive_1980, barnes_formation_2002}. This makes them excellent multimessenger sources as they emit low-frequency gravitational waves (GW) during their inspiral~\cite{peters_gravitational_1964} and electromagnetic (EM) radiation from accretion and outflows~\cite[see, e.g.,][and references therein]{bogdanovic_electromagnetic_2022}. Their multimessenger nature also makes observations of SMBBHs very important, as they can help probe fundamental astrophysics, gravity and cosmology~\cite{LISA_white_paper, lisa_gw_whitepaper, lisa_multimessenger_whitepaper, arun_new_2022, amaro-seoane_astrophysics_2023}.

GWs emitted by SMBBHs are primary targets for both the Pulsar Timing Array (PTA) and the future Laser Interferometer Space Antenna (LISA), albeit at different frequencies~\cite{LISA_white_paper, lisa_gw_whitepaper}. PTAs have reported evidence for a stochastic background GW~\cite{agazie_nanograv_2023, nanograv} but have yet to resolve individual SMBBHs~\cite{rosado_expected_2015, kelley_single_2018, aggarwal_nanograv_2019}, and LISA is expected to launch in the mid-2030s~\cite{lisa_paper, amaro-seoane_astrophysics_2023, lisa_consortium_waveform_working_group_waveform_2023, arun_new_2022}. On the other hand, EM surveys have revealed 100s of SMBBH candidates~\cite{rodriguez_2006, rodriguez_2009, charisi_multiple_2015, graham_systematic_2015, panstars_smbh, oneill_unanticipated_2022, kiehlmann_pks_2024}. However, without a horizon-scale observation with, e.g., the Event Horizon Telescope (EHT), or a direct GW detection, we cannot confirm these candidates as SMBBHs. Therefore, to help interpret EM observations of SMBBH candidates, and to identify EM counterparts to GWs of SMBBHs, the field has turned to predicting `smoking gun' EM signals from SMBBHs with numerical and theoretical methods.

Recent horizon-scale observations by the Event Horizon Telescope (EHT) suggest that both M87* and Sag A* host magnetically arrested disks (MADs)~\cite{eht_m87,eht_saga_2022}. MADs form when poloidal magnetic field saturates on the black hole (BH) horizon and becomes dynamically important, thereby regulating the accretion of matter~\cite{ narayan_mad_2003}. MAD  efficiencies ($P_{\infty}/\dot Mc^2$, with $P_{\infty}$ the power carried out to infinity and $\dot M$ the rest-mass accretion rate) can be up to $50\%$ for nonspinning BHs~\cite{igumenshchev_mad_2003, narayan_mad_2003} and potentially greater than 100\% for spinning BHs~\cite{tchekhovskoy_mad_2011}. This makes MADs around BHs prime candidates for being the engine that powers active galactic nuclei (AGN), galactic-scale relativistic jets, jetted tidal disruption events, and gamma-ray bursts ~\cite{tchekhovskoy_swift_2014, eht_m87, lalakos_bridging_2022, gottlieb_collapsar_2023}.

As a result, MADs around single BHs have been studied extensively using general-relativistic magnetohydrodynamic (GRMHD) simulations initialized with a torus~\cite{chakrabarti_1985} threaded with an initially weak poloidal magnetic field~\cite{tchekhovskoy_mad_2011, mckinney_mad_2012}. These studies revealed a few key features: 1. saturation and fluctuation of magnetic flux on the horizon; 2. ejection of magnetic flux from near the horizon; 3. launching of highly efficient jets that extract BH angular momentum~\cite{lowell_rapid_2023}; and 4. accretion proceeding through thin sheets at the equator, and the presence of current sheets. Simulations of MADs around nonspinning BHs showed similar features, except for the relativistic jet, and also added that flux eruptions can drive outward angular momentum transport~\cite{chatterjee_flux_2022}. Additionally, the MAD end-state appears to be largely agnostic to initial conditions: recent work has shown that disks threaded with toroidal magnetic fields~\cite{liska_large-scale_2020, begelman_what_2022,jacquemin-ide_magnetorotational_2024, manikantan_winds_2024} and spherical Bondi-like accretion flows~\cite{lalakos_bridging_2022,gottlieb_black_2022,gottlieb_collapsar_2023, lalakos_jets_2024} can both give rise to MADs. 

MADs also have significant implications for observable transients from galactic nuclei. Horizon-scale studies of magnetic reconnection in MADs suggest that magnetic flux eruptions can power the observed infrared and X-ray flares from Sag A*~\cite{dexter_sgr_2020, ripperda_magnetic_2020, scepi_sgr_2022, ripperda_black_2022}. Therefore, if MAD conditions were to arise in the context of BBH accretion, such flaring activity could produce EM counterparts to their GWs and would add a new paradigm to our understanding of emission from SMBBHs. However, it is unclear if the minidisks that form around each BH in a binary can become magnetically arrested. Around single BHs, the accretion flow typically reaches an approximately axisymmetric steady state that allows it to continuously accumulate magnetic flux on the horizon. In the case of (near-equal mass) BBHs accreting from a circumbinary disk (CBD), the minidisks go through phases of replenishment and depletion as matter is accreted onto each component through quasiperiodic accretion streams. As a result, each binary component does not host a roughly axisymmetric, steady state flow. Under these conditions, a natural question arises: can minidisks in circumbinary flows become MAD?

So far, MADs have received limited attention in the context of SMBBHs~\cite{most_magnetically_2024, most_decoupling_2025, ressler_dual_2025, wang_bmad_2025}. Recent Newtonian MHD simulations of CBD accretion suggest the possibility of a magnetically arrested cavity~\cite{most_magnetically_2024, most_decoupling_2025, wang_bmad_2025}. However, to understand horizon-scale processes such as magnetic flux saturation on the BH horizons, ejection of magnetic flux bundles, magnetic reconnection, and jet launching, general-relativistic simulations are necessary~\cite{tchekhovskoy_mad_2011, mckinney_mad_2012, liska_large-scale_2020, gottlieb_black_2022, ripperda_magnetic_2020, ripperda_black_2022}. 

Some general-relativistic approaches make use of approximate spacetimes~\cite{combi_superposed_2021, combi_binary_2024}, which are effective at modeling horizon-scale dynamics at large binary separations~\cite{combi_superposed_2021, lopez_armengol_circumbinary_2021, combi_binary_2024, avara_accretion_2024}. Recent work at binary separations of $d/M \lesssim 27$ have revealed MAD-like features from Bondi-like initial conditions~\cite{ressler_dual_2025}. Bondi-like conditions allow for continuous feeding of the BHs, as opposed to quasiperiodic accretion from accretion streams. This, in-principle, allows the BHs to accumulate more magnetic flux on their horizons. The leading AGN accretion model, however, requires accretion from a rotationally supported disk. Therefore, the question remains: can a MAD state arise from CBD accretion onto BBHs?

Reliably modeling SMBBH accretion from first-principles during the late inspiral and eventual merger requires simulations in full GR, where Einstein's equations are solved in addition to the MHD equations~\cite{duez_relativistic_2005, farris_binary_2011, farris_binary_2012, gold_accretion_2014, khan_disks_2018, gold_relativistic_2019, paschalidis_minidisk_2021,cattorini_misaligned_2022, bright_minidisk_2023, cattorini_grmhd_2024, ruiz_general_2023, fedrigo_grmhd_2024, manikantan_coincident_2025, manikantan_effects_2025, ennoggi_relativistic_2025, ennoggi_merger_2025}. Simulations in full GR are necessary to determine if minidisks around the individual BHs can become MAD, as we need to accurately capture the horizon-scale interactions of the magnetized fluid and the BHs.

In this work, we simulate the MHD accretion flow of a magnetized circumbinary torus around a BBH at initial orbital separation $d/M = 30$, in full $3+1$ GR. We discover that the minidisks become MAD, and describe their features, including the dimensionless horizon flux, horizon-scale dynamics, magnetic flux eruptions, and the evolution of ejected magnetic flux. 

The paper is structured as follows. In Section~\ref{sec:methods} we discuss the initial data and numerical evolution of our binary. In Section~\ref{sec:results} we outline the key features of our magnetically arrested BBH accretion flow, and in Section~\ref{sec:summary} we summarize and discuss our findings. In the Appendix we outline the diagnostics used, including how we evaluate magnetic flux on the horizons. 

Throughout, we adopt geometrized units in which $G=c=1$, where $G$ is the gravitational constant and $c$ is the speed of light. Our spatial and time domains are measured in units of $M$, where $M$ is the Arnowitt-Deser-Misner (ADM) mass of the spacetime. Therefore, in the spatial domain $1M = GM/c^2$ and in the time domain $1M = GM/c^3$. 


\section{Methods} \label{sec:methods}
    We perform our simulations within the \einstein{} framework~\cite{einsteintoolkit}. We outline our diagnostics, including our new \einstein{} thorn for measuring the magnetic flux on each horizon, in the Appendix.
    
\subsection{Initial Data}
    \paragraph{Spacetime.}
        We use the \texttt{TwoPunctures} code to prepare the spacetime initial data~\cite{twopunctures_ansorg, twopunctures_paschalidis}. We initialize the BHs with no spin ($\chi_1 = \chi_2 = 0$), equal mass ($q = m_2/m_1 = 1$), and on a quasicircular orbit with separation $d/M = 30$, which, to our knowledge, is the largest separation to date studied in full $3+1$ general relativity.
    \paragraph{Grid.} 
        We use a three-dimensional cartesian grid with the outer boundary extending from $-5120 \, M$ to $+5120 \, M$ in the $x, y$ and $z$ directions. We use \texttt{Carpet} to employ a box-in-box adaptive mesh refinement (AMR) scheme~\cite{schnetter_carpet_2016}, with a total of 14 refinement levels. We use 3 sets of nested refinement levels: one centered on each BH and the third on the origin, which is also the location of the binary center of mass. The half-side length of each refinement level is $5120 \times 2^{1-i} \, M, i = 1,\dots 14$ and the resolution at the finest (coarsest) refinement level is $\Delta x = M/64$ ($\Delta x = 128 \, M$). This resolution is motivated by a convergence study in~\cite{manikantan_effects_2025}, and was also shown to be sufficient for spacetime evolutions of equal mass, nonspining black holes at even larger orbital separation of 100$M$~\cite{Lousto:2013oza}. Additionally, we set the resolution such that we resolve the fastest growing MRI wavelength with at least 20 cells in the bulk of the disk and 35 cells near the inner edge of the disk.
        
    \paragraph{Matter.}
        We use the power-law torus solution for our CBD initial conditions~\cite{chakrabarti_1985, villiers_magnetically_2003, khan_disks_2018}. We set the inner edge of the torus at $r=27M$ with specific angular momentum $l=6.06$. These two parameters along with the equation of state determine the outer edge of the CBD at $r \simeq 130M$. We use a $\Gamma-$law equation of state with $\Gamma=4/3$, which is  appropriate for radiation pressure dominated accretion flows.
    \paragraph{Magnetic Fields.}
        To trigger the magnetorotational instability (MRI)~\cite{Balbus},
        we initialize our disk with a seed poloidal magnetic field as in~\cite{khan_disks_2018} with minimum initial plasma beta $\beta \equiv P_{\rm gas}/P_{\rm mag} \simeq 26$. This value is similar to that of other studies of magnetized accretion onto binaries~\cite{most_magnetically_2024}. The initial maximum magnetization in our disk ($\sigma \equiv b^2/8\pi\rho_0$) is $\sigma \sim 1.11 \times 10^{-5}$, where $b$ is the magnitude of the magnetic field measured by an observer comoving with the plasma. 
\subsection{Evolution}
    \paragraph{Spacetime.}
        We evolve the spacetime by solving the Einstein equations in the Baumgarte-Shapiro-Shibata-Nakamura (BSSN) formalism~\cite{Shibata1995, Baumgarte1998}, as implemented in the \texttt{LeanBSSN} thorn using 6th-order finite differences~\cite{sperhake_binary_2007}. We adopt the moving puncture gauge conditions~\cite{baker_gravitational-wave_2006, campanelli_accurate_2006} with the shift vector parameter $\eta$ set to $\eta = 1.405/M$.
    \paragraph{Magnetohydrodynamics.} 
        We employ the $3+1$ general-relativistic MHD AMR code \ilgrmhd{} within the \einstein~\cite{ilgrmhd}. \ilgrmhd{} evolves the equations of ideal MHD in flux-conservative form via the Harten-Lax-van Leer Riemann solver~\cite{toro_riemann_2009}, and the Piecewise-Parabolic method for reconstruction~\cite{colella_piecewise_1984}. We adopt the generalized Lorenz gauge condition from~\cite{farris_binary_2012} for our EM gauge, and set the Lorenz gauge damping parameter to $\xi = 8/M$ to remove spurious gauge modes~\cite{etienne_em_gauge}. We do not treat radiative feedback, heating, or cooling. Finally, the fluid does not backreact onto the spacetime as the spacetime's mass/energy content is dominated by the SMBBH.
    
\begin{figure*}[ht]
        \centering
        \includegraphics[width=1\textwidth]{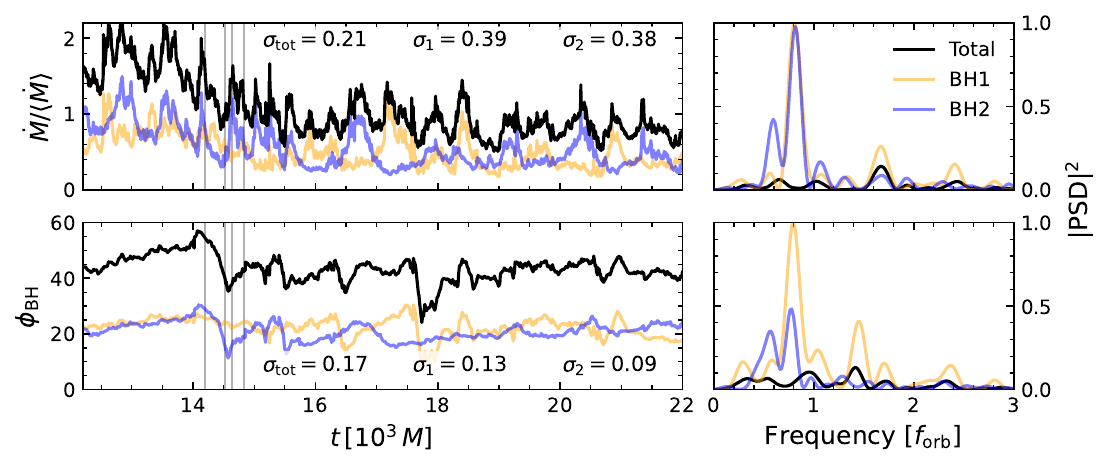}
        \caption{Top: Rest-mass accretion rate $\dot{M}$ onto each BH (orange and blue lines) and the sum of the two (solid black line) normalized by the total time averaged $\langle \dot{M} \rangle$ for the time period $15 \lesssim t/10^3 M \lesssim 22$. Bottom: Dimensionless magnetic flux $\phi$ on each BH (orange and blue lines) and the sum of the two (solid black line). We also include the standard deviation of each timeseries. The vertical translucent gray lines indicate times plotted in Figure~\ref{fig:single}. Right: Normalized power spectral density (PSD) of the Fourier transforms of the signals in the left column. We normalize the frequency axis by the binary's orbital frequency and normalize the amplitudes by the maximum of the three PSDs shown. The individual signals all show a peak periodicity at $f\sim 0.8 \, f_{\rm orb}$, whereas the total signals show a suppressed variability with the $\dot{M}$ peaking at $f\sim 1.6 \text{ and } 1.0 f_{\rm orb}$, while $\phi_{\rm BH}$ peaks at $f\sim 1 \text{ and } 1.5 \, f_{\rm orb}$. We have checked that the Fourier transform of the unnormalized horizon flux is the same.}
        \label{fig:phi}
    \end{figure*} 

\section{Results} \label{sec:results}
    The initial evolution of our accreting BBH is qualitatively similar to previous studies of CBDs in full general relativity~\cite{duez_relativistic_2005, farris_binary_2011, farris_binary_2012, gold_accretion_2014, khan_disks_2018, gold_relativistic_2019, paschalidis_minidisk_2021,cattorini_misaligned_2022, bright_minidisk_2023, cattorini_grmhd_2024, ruiz_general_2023, fedrigo_grmhd_2024, manikantan_coincident_2025, manikantan_effects_2025, ennoggi_relativistic_2025, ennoggi_merger_2025}. The binary torques the CBD inner edge to form accretion streams that circularize around the BHs to create minidisks, as well as filling the cavity with matter~\cite{paschalidis_minidisk_2021, combi_minidisk_2022, bright_minidisk_2023}. Since our binary is initially at separation $d \simeq 30M$, they have large spheres of influence that allow them to form larger minidisks than previously seen in full GR. Here, we describe the key characteristics of the BBH accretion flow to demonstrate that the minidisks become magnetically arrested: 1. the dimensionless magnetic flux on each horizon, 2. the horizon-scale matter flow and magnetic field structure, and 3. the ejection of magnetic flux from the horizon vicinity. We perform the majority of our analysis for the time period $t \sim 14,000 - 22,000 \, \rm M$, where the binary has inspiraled to a separation of $d\sim 29-28 \, \rm M$. These separations are larger than the expect decoupling radius for this binary, which we estimate to be $13$M using Eq. 5 of~\cite{Gold:2013zma}.
    
    \subsection{Dimensionless Horizon Fluxes} 
    After about 14 orbits ($t \simeq 15000M$), the rest-mass accretion rate settles into a quasisteady state -- its moving average  remains approximately constant -- and each horizon is saturated with magnetic flux. In the top row of Fig.~\ref{fig:phi}, we report the rest-mass accretion rate onto each BH (orange and blue lines) normalized by the time average of the total rest-mass accretion rate (solid black line). In the bottom row we report the dimensionless magnetic flux (Eq.~\eqref{eq:dimensionless_flux}) on each horizon and their sum. In the left column we report the timeseries and in the right we report the normalized power spectral density (PSD) of their Fourier transform. We normalize each PSD to the maximum of the three shown in the panel. We do this so that the peaks of each PSD are more easily distinguishable, not to compare their amplitudes. We wait until after 14 orbits ($\sim 15000 \, M$) to analyze the system -- when the accretion rate has settled.


    The total rest-mass accretion rate (top left panel) exhibits variability that, after $t \simeq 16000M$, is driven by alternating accretion episodes onto the individual BHs. 
    The power spectral density (PSD) of the Fourier transform of the rest-mass accretion rate (top right panel) reveals a dominant periodicity at $\sim 0.8 \, f_{\rm orb}$ for each individual BH (orange and blue lines), where $f_{\rm orb}$ is the orbital frequency of the binary determined via $\ell=m=2$ mode of the gravitational waves we compute. The PSD of the total accretion rate (black line) exhibits dominant periodicity at $1.6 \, f_{\rm orb}$, and secondary peak at $1 \text{ and } 0.6\, f_{\rm orb}$. Previous studies of CBD accretion onto quasicircular binaries, both in 2D and 3D~\cite{paschalidis_minidisk_2021, westernacher-schneider_multi-band_2022, bright_minidisk_2023, manikantan_effects_2025, ennoggi_relativistic_2025}, often find $1.4 \, f_{\rm orb}$ for the periodicity of the total rest-mass accretion rate of quasicircular binaries, which is close to the dominant $1.6 \, f_{\rm orb}$ peak we find here. 
    
    It is unclear why the $1.4 \, f_{\rm orb}$ periodicity is not present in our simulations. It is possible that the dominant periodicity of the accretion rate at relativistic separations is dependent on the binary separation, and in this study we probe separations that are 50\% larger than any previous simulations in full GR. It is also possible that the MAD state of our minidisks affects the variability. Furthermore, the alternating accretion pattern suppresses the variability of the total accretion rate, as demonstrated by the standard deviation of the accretion rates shown in the top left panel. The standard deviation of the total rest-mass accretion rate ($\sigma_{\rm tot} = 0.21$) is about half that of the individual rest-mass accretion rates ($\sigma_1 = 0.39; \sigma_2 = 0.38$).

    The dimensionless horizon magnetic flux timeseries (bottom left panel) are characterized by an initial steady increase on each BH to a value $\phi_{\rm BH}\sim 25-30$ followed by rapid drops. The rapid drops correspond to ejection of magnetic flux from the horizon vicinity -- often also referred to as magnetic flux eruption events or flares. This is a tell-tale characteristic of MADs~\cite{tchekhovskoy_mad_2011}. There are a few eruption events we can identify in both timeseries -- on BH1 we can identify them at $t \simeq 16400M$ and $17700M$ and for BH2 we can identify them at $t \simeq 14400M$ and $15400M$. They seem to arise every about $1000M$, which is very close to the binary orbital period.

    Additionally, the Fourier transform of the individual $\phi_{\rm BH}$ signals (bottom right panel) also exhibits a periodicity at $\sim 0.8 \, f_{\rm orb}$ which is the same as the rest-mass accretion rate onto the individual black holes. The PSD of the sum of the horizon fluxes reveals a periodicity of $\sim 1.4 \text{ and } 0.95 \, f_{\rm orb}$ for the range $15,000 < t/M < 22,000$. And, contrary to the total rest-mass accretion rate, the standard deviation of the sum of the horizon fluxes ($\sigma = 0.17$) is greater than the standard deviation of the individual signals ($\sigma_1 = 0.13; \sigma_2 = 0.09$).

    It is not clear what physical mechanism may be driving the difference in the standard deviation (variability) of the mass accretion rate and the horizon fluxes. It may not be surprising that the accretion rate variability is not entirely reflected on the variability of the normalized horizon magnetic flux-- in the case of single BH accretion the magnetic flux time series is much smoother than the accretion rate time series~\cite[see, e.g., ][]{tchekhovskoy_mad_2011}. Upon closer inspection of the timeseries, the  larger standard deviation in the summed horizon fluxes arises because the sum increases the deviations from the individual horizon flux averages. By contrast, the individual mass accretion rates are about $\pi$ out of phase which means that the sum of the timeseries works to reduce the deviation from the average.

    \begin{figure*}[ht]
        \centering
        \includegraphics[width=\textwidth]{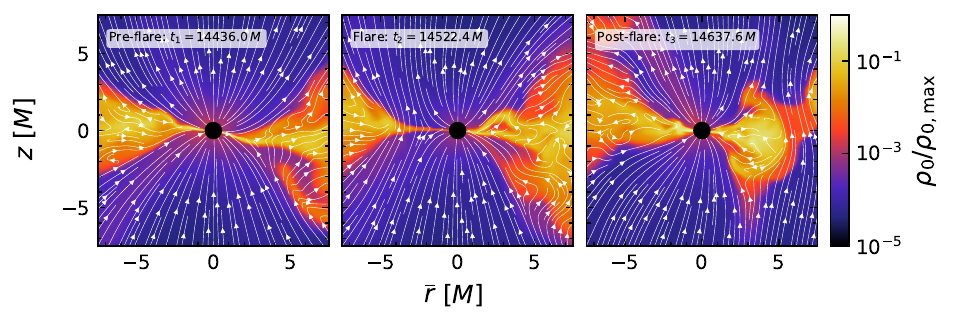}
        \caption{We plot the rest-mass density and magnetic field structure at horizon-scales at a pre-flare quiescent state (left panel), during an active flaring state (center panel), and at a post-flare quiescent state (right panel). The rest-mass density $\rho_0$ is normalized by its initial maximum in the disk $\rho_{0, \rm max}$ on a color scale where brighter colors indicate higher densities. We overplot the magnetic field with directed streamlines and indicate the BH horizon with a black disk. In the left panel, accretion proceeds through an equatorial flow that is slightly displaced from the binary orbital plane. In the center panel, there is an active flux eruption to the left of the BH, where the vertical magnetic field lines are ordered, thread the binary orbital plane, and extend vertically, and the density has dropped to $\rho_0/\rho_{0, \rm max}\sim 10^{-2}$. In the right panel, the flux bundle is at $\bar r \sim6M$ on the right hand side of the minidisk and shares the same magnetic field characteristics as in the middle panel.}
        \label{fig:horizon}
    \end{figure*}
    
    \subsection{Horizon Scale Magnetic Field} 
    
    Magnetic flux eruptions are driven by poloidal magnetic fields, which thread the horizon, squeezing at the midplane, reconnecting, and being ejected outwards. In Figure~\ref{fig:horizon}, we plot the horizon-scale rest-mass density and magnetic field structure at three characteristic states: 1. pre-flare quiescent ($t/M = 14436.0$), 2. flaring ($t/M = 14522.4$; flare is to the left of the BH where the magnetic field outside the BH is ordered and the density is low), and 3. post-flare quiescent ($t/M = 14637.6$). We plot the rest-mass density ($\rho_0$) normalized by its initial maximum ($\rho_{0, \rm max}$) with a color map. We also plot the magnetic field measured by a normal observer with directed white streamlines and indicate the BH horizon with a black disk. Throughout we use overline notation, e.g. $\bar{r}$, to indicate a coordinate centered on a single BH. 

    In the left panel, the horizon is saturated with magnetic flux ($\phi_{\rm BH} \simeq 30$; see Fig.~\ref{fig:phi}). The minidisk is at the equator with a thickness $H/r \sim 0.3$ and density $\rho_0/\rho_{0, \rm max} \sim 10^{-1}$. Below and above the BH are low density regions ($\rho_0/\rho_{0, \rm max} \sim 10^{-4}$) where matter has been evacuated by the polar jet outflows. The incipient jet region is defined by the highly ordered paraboloidal magnetic field lines that thread the horizon, extend radially, and then collimate in the $z-$direction. The accretion flow is squeezed at the equator by the incipient jet above and below the BH. As the flow approaches the BH at the equator, it proceeds through two thin, higher density sheets ($\rho_0/\rho_{0, \rm max} \sim 1$), which also host current sheets as exhibited by the magnetic field lines above and below the equator.

    In the center panel, we plot the system during an active flare. To the left of the BH, we see that the density has dropped by an order of magnitude at the equator ($\rho_0/\rho_{0, \rm max} \sim 10^{-2}$). At higher resolutions, this may be even more drastic~\cite{ripperda_black_2022}. A bundle of large-scale, vertical magnetic field lines on the left of the BH thread the midplane. Towards the outer edge ($\bar{r} \simeq -4$) the field lines are almost vertical while closer to the horizon ($\bar{r} \simeq -2$) the field lines have an equatorial Y-point, suggesting magnetic reconnection took place through the equatorial current sheet. There are still vertical field lines threading the BH horizon and the flow to the right of the horizon shares similarities to the pre-flare quiescent state. 

    Finally, in the right panel, the accretion flow returns to a post-flare quiescent state, where accretion proceeds through thin sheets at the equator with current sheets beginning to form. The flux eruption is still making its way out of the minidisk, now on the right of the horizon at $\bar{r}\sim 6$. This horizon-scale behavior is consistent with studies of accretion onto single BHs and, therefore, we expect many of the observational implications, such as infrared and X-ray flares due to magnetic reconnection and plasmoids, to carry over to the binary case~\cite{dexter_sgr_2020, ripperda_magnetic_2020, scepi_sgr_2022, ripperda_black_2022}.

    \begin{figure*}[ht]
        \centering
        \includegraphics[width=\textwidth]{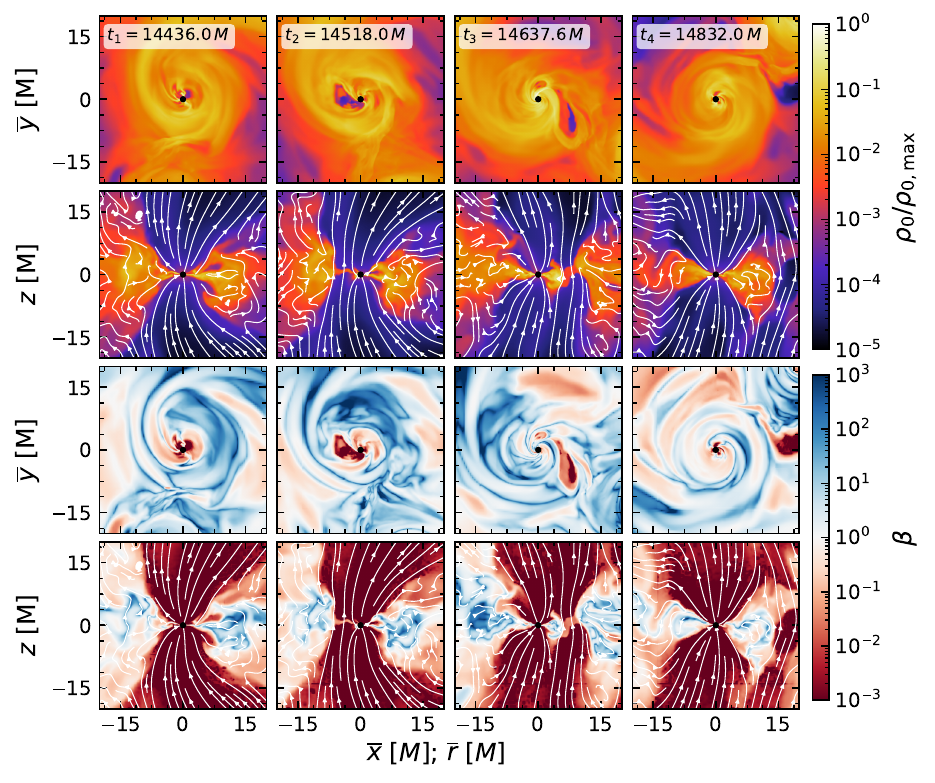}
        \caption{Top two rows: slices of the rest-mass density ($\rho_0$) normalized by its initial maximum ($\rho_{0, \rm max}$) on a color scale where brighter colors indicate higher densities. Bottom two rows: slices of the plasma beta ($\beta \equiv P_{\rm gas}/P_{\rm mag}$) on a color scale where white indicates equipartition and red (blue) indicates that magnetic (gas) pressure dominates. We overplot magnetic field lines in white and indicate the BH horizon with a black disk. First and third rows: we plot the system in the orbital plane of the binary. Second and fourth rows: we plot the system in the $\bar{r}-z$ plane, with $\bar{r}$ centered around the BH and going through a magnetic flux bundle. The first column plots the system when the horizon magnetic flux is saturated before an eruption event. The following three columns show the magnetic flux bundle erupting from the horizon, moving through the minidisk, and entering the cavity.}
        \label{fig:single}
    \end{figure*}
    
    \subsection{Magnetic Flux Eruptions} \label{subsec:eruption}
    
    A key feature of MADs is the ejection of bundles of large-scale, vertical, magnetic fields from the horizon vicinity~\cite{tchekhovskoy_mad_2011, begelman_what_2022}. In Figure~\ref{fig:single}, we outline the evolution of such a magnetic flux eruption event. 

    In the top two rows of Fig.~\ref{fig:single} we plot rest-mass density of the fluid normalized to its initial maximum value ($\rho_0/\rho_{0, \rm max}$) with a color map. In the first (second) row we show $\bar{x}-\bar{y}$ ($\bar{r}-z$) slices and also overplot the magnetic field measured by a normal observer with directed white lines. In the bottom two rows we plot the plasma beta ($\beta \equiv P_{\rm gas}/P_{\rm mag}$) on diverging color maps. Each row describes the same slices as the top two rows. 
    
    In the left most column, at $t = 14436.0M$, we show the system in a quiescent state, where there is no active magnetically driven eruption. A persistent minidisk of density $\rho_0/\rho_{0, \rm max} \sim 10^{-1}$ has formed around the BH with radius $\bar{r}\sim 12 \, M$ (first row). Vertical magnetic field lines thread the horizon and the accretion flow is squeezed into sheets at the equatorial plane (second row, see also Figure~\ref{fig:horizon}). The plasma beta within the disk is gas pressure dominated ($\beta>1$), however near the horizon ($\bar{r} \lesssim 2$) magnetic pressure dominates (third row). The high plasma beta value near the horizon is due to the accretion flow being displaced from the equator (see left panel in Figure~\ref{fig:horizon} where the accretion stream is below the $x=0$ plane) and, as a result, our $x$-$y$ slice is cutting through part of the incipient jet which is magnetic pressure dominated. Furthermore, outside the minidisk, the low-density cavity approaches equipartition ($\beta \sim 1$). The vertical slice shows that the polar regions of the BH are magnetic pressure dominated (fourth row). 
    
    In the second column, at $t = 14518.0M$, we plot the system as a magnetic flux bundle is ejected from the vicinity of the BH horizon. This bundle is characterized by a much lower density ($\rho_0/\rho_{0, \rm max} \sim 10^{-4}$) and much lower fluid plasma beta ($\beta \lesssim 10^{-2}$). Furthermore, in the vertical slices, this low-density, low $\beta$ region is carved out by a bundle of ordered vertical magnetic fields that thread the equatorial plane. In the third and fourth columns, after completing half an orbit, the magnetic flare appears on the right of the horizon. In the third column ($t = 14637.6M$), the flux bundle is within the minidisk at $\bar{r}/M \simeq 6$ and is still characterized by low-density and $\beta$. In the fourth column ($t = 14832.0M$), the flux bundle is at $\bar{r}/M \simeq 15$, and has left the minidisk and entered the lower-density ($\rho_0/\rho_{0, \rm max} \sim 10^{-3}$) cavity. The flow on the BH horizon returns to its pre-flare state, albeit with a lower magnetic flux threading the horizon (see Figure~\ref{fig:phi}). The minidisk now appears to be in equipartition, $\beta\sim1$, which is consistent with previous MAD literature~\cite{liska_large-scale_2020, jacquemin-ide_magnetorotational_2024}.

    These magnetic flares can play a significant role in the outward transport of angular momentum~\cite{chatterjee_flux_2022, most_magnetically_2024}. Furthermore, large-scale magnetic fields can lead to winds and ejecta that torque the accretion disk~\cite{manikantan_winds_2024}. A torque analysis as in~\cite{chatterjee_flux_2022, most_magnetically_2024, manikantan_winds_2024} is not gauge invariant in the BBH case because there exists no axial killing vector. Therefore, we do not perform such an analysis here.

\section{Summary and Discussion} \label{sec:summary}

In this work we have performed MHD simulations in full GR that reveal the formation of magnetically arrested minidisks (MAM) around binary black holes. We simulate $\sim 23000 \, M$ ($\sim 22$ orbits), which, to our knowledge, is the longest MHD simulation of CBD binary accretion in full $3+1$ GR, and at the largest initial orbital separation, $d\sim 30M$. 

We summarize our key findings here:

\begin{enumerate}[leftmargin=1em,labelwidth=*,noitemsep]
    \item The dimensionless magnetic flux on each horizon saturates at $\phi_{\rm BH} \sim 25-30$.

    \item The variability of the total rest-mass accretion rate is suppressed by alternating accretion onto the BHs but still exhibits dominant periodicities with peaks in the PSD at $f \sim 1.6$ and $1\, f_{\rm orb}$. The total $\phi_{\rm BH}$ shows relatively weak variability on the orbital period ($f\sim 1 \text{ and } 1.4\, f_{\rm orb}$). 
    
    \item Accretion near the horizon $\bar r/M < 5$ proceeds through thin accretion sheets that are squeezed into the equatorial plane by the poloidal magnetic flux threading the horizon. Additionally, we observe the formation of current sheets on horizon-scales.

    \item Magnetic reconnection takes place, and bundles of vertical magnetic fields are ejected from the horizon vicinity, travel through the minidisk, and populate the cavity. Our cavity is not magnetically arrested and tidal streams proceed unperturbed by the vertical magnetic field.
\end{enumerate}

Qualitatively, the key features we list above are smoking-gun evidence that our BBHs host MAMs. We note that our measured dimensionless horizon fluxes ($\phi_{\rm BH} \sim 30$) are lower than that typically found in single BH MADs ($\phi_{\rm BH}\sim 50-80$)~\cite{tchekhovskoy_mad_2011, mckinney_mad_2012, liska_large-scale_2020, begelman_what_2022, lalakos_jets_2024}. However, there is no universal value of the saturation value of $\phi_{\rm BH}$; the measured value depends on feedback mechanisms such as jet and flux tube ejection (see also Appendix A of~\cite{ressler_dual_2025})\footnote{Additionally, recent work suggests that the horizon flux value is both equation-of-state and BH spin dependent - lower $\Gamma$ values and lower spins both lead to lower horizon flux values; Personal communication with Lalakos et al.}. Furthermore, the accretion flow onto a BBH is very different than that onto a single BH. For example, the flow onto a BBH and its minidisks is modulated by the periodicity of the streams that are pulled from the inner edge of the CBD. This naturally regulates how much magnetic flux can reach and persist on the BH horizons. Moreover, the geometry of the minidisks around each BH tends to be somewhat eccentric -- it is unknown what effect this would have on the horizon flux value. Despite these key differences, our BBH exhibits the tell-tale features of MADs: magnetically driven eruptions, current sheets at the horizon, and saturation of magnetic flux on the horizon. 

Some caveats are in order: our simulations are resolution limited. While the resolution we adopt ($M/64$ at the highest refinement level) is typical for BBH accretion in full $3+1$ GRMHD~\cite{paschalidis_minidisk_2021, bright_minidisk_2023, manikantan_coincident_2025, manikantan_effects_2025, ennoggi_relativistic_2025}, simulations at higher resolutions demonstrate significant differences in the accretion flow at the horizon-scale. For example, low-resolution simulations of MADs onto single BHs tend form thicker current sheets and have a higher magnetic field reconnection rate~\cite{ripperda_black_2022}. However, the qualitative picture of MADs remains invariant with resolution. It is possible that limited resolution affects some quantitative features of our simulations. Additionally, we have only simulated an initially purely poloidal magnetic field geometry in this work, which, at least in the case of single BHs, is more conducive to forming MADs as the large-scale vertical flux is seeded in the CBD. However, recent studies in single BH accretion have shown that the MAD end-state can arise from a variety of initial magnetic field and gas density conditions~\cite{liska_large-scale_2020, begelman_what_2022,jacquemin-ide_magnetorotational_2024, lalakos_bridging_2022, gottlieb_black_2022, gottlieb_collapsar_2023, lalakos_jets_2024}. Therefore, we expect that MAMs in BBH accretion can arise from a variety of magnetic field initial conditions, but further  simulations are necessary to establish this expectation.

The presence of MAMs, and the magnetic flux eruptions they produce, may have significant implications for observations of SMBBHs. For example, reference~\cite{ripperda_black_2022} found that plasmoid-mediated magnetic reconnection can occur near the horizon. These plasmoids can accelerate electrons to nonthermal energy distributions and even power X-ray flares~\cite{ripperda_magnetic_2020}. We expect that these EM signatures will translate over to circumbinary accretion. To test this, we performed an emissivity estimate of our system as described in equation 18 of~\cite{porth_event_2019} and found that the lightcurve of the optically thin thermal synchrotron emission follows the dimensionless horizon flux timeseries closely. Meaning, drops in horizon flux corresponded directly to drops in luminosity. This is not surprising as the emission proxy from~\cite{porth_event_2019} is calculated directly from primitive variables that are affected by magnetic eruptions, namely, pressure, density, and the magnetic energy. This estimate did not show a clear periodicity in the lightcurve. However, this estimate does not account for flaring from magnetic reconnection and eruptions. 

We observed magnetic flux eruptions occurring on the orbital period for $\sim4$ orbits, which might suggest quasiperiodicity in flaring. We have also observed eruptions in simulations of quasicircular and eccentric BBH accretion at smaller separations~\cite{manikantan_coincident_2025, manikantan_effects_2025}. Importantly, we noticed a rapid increase in the frequency of magnetic eruptions as the binaries approached merger (see supplemental material for a movie of the simulation from~\cite{manikantan_effects_2025}). This suggests a connection to the binary orbital period. Regardless, additional studies are necessary to evaluate the connection of the flux eruption frequency to the binary orbital frequency. Moreover, a full general-relativistic ray-tracing (GRRT) and radiative transfer is necessary to evaluate observable EM emission from this binary and its MAMs.

Increased flaring approaching merger would have significant implications for EM observations of these systems. As the binary approaches merger its inspiral outpaces the inward motion of the CBD inner edge, leading to the so-called binary-disk decoupling. During decoupling, the rest-mass accretion rate drops and EM signals powered by accretion are expected to weaken. After merger, the disk fills in the cavity left behind by the binary, begins to accrete onto the remnant, and eventually repowers bright EM emission -- the so-called rebrightening. However, our simulations demonstrate that magnetic flux eruptions occur even during the decoupling phase, and at an increasing rate, which may lead to repeated flaring in the X-ray and IR through merger. Again, a more complete analysis with GRRT and radiative transfer would be necessary to evaluate this.

The increased frequency of flux eruptions close to merger can be explained by a few mechanisms. The mechanism in our simulations correlates with binary-disk decoupling. As the binary decouples from the CBD, the ambient pressure in the cavity and the ram pressure of the matter accreting onto the horizons both drop. This allows magnetic pressure to more easily balance and overcome the lower ambient pressure environment. This is consistent with what we observe in our simulations: the black holes leave a lower density/lower pressure environment ``behind" them as they sweep the inner cavity material, which does not have enough time to refill as they move faster, leading to the magnetic flux eruptions taking place ``behind" the black holes.

While there is recent work reporting magnetically arrested accretion flows in the context of cloud (Bondi-like) accretion in BBH spacetimes~\cite{ressler_dual_2025}, our work demonstrates MAD flow properties in CBD accretion for the first time. There is a lot to still be established. For example, longer-term evolutions of CBDs are needed. Additionally, the flow features with and without black hole spin may differ. For example, spinning black holes can have smaller innermost stable orbits, which allow even larger minidisks to form~\cite{bright_minidisk_2023}. Furthermore, varying the binary mass ratio may also complicate this picture as it would affect the size of the minidisks and the relative accretion rates between the binary components~\cite{gold_accretion_2014}. We plan to continue evolving our existing simulation and to probe these other parts of the parameter space in future work. Furthermore, we will perform a general-relativistic ray-tracing of the system to establish the effect of these flux eruptions on EM signatures from SMBBHs.

\begin{acknowledgments}
We are grateful to Aretaios Lalakos, Elias Most, Ziri Younsi, and Chi-Kwan Chan for insightful discussions during the preparation of this manuscript. This work was in part supported by NASA grant 80NSSC24K0771 and NSF grant PHY-2145421 to the University of Arizona. This research is part of the Frontera computing project at the Texas Advanced Computing Center. Frontera is made possible by U.S. National Science Foundation award OAC-1818253. This work used Stampede2 and Stampede3 at the Texas Advanced Computing Center through allocation PHY190020 from the Advanced Cyberinfrastructure Coordination Ecosystem: Services \& Support (ACCESS) program, which is supported by U.S. National Science Foundation grants 2138259, 2138286, 2138307, 2137603, and 2138296~\citep{boerner_access_2023}.
\end{acknowledgments}

\appendix \label{appendix}
\section{Diagnostics and Horizon Magnetic Flux}

    We use \texttt{Kuibit} for our analysis~\cite{kuibit}. We use the methods outlined in~\cite{farris_binary_2012, bright_minidisk_2023} to measure the rest-mass accretion rate onto each horizon and the Hill sphere mass, locate apparent horizons with \texttt{AHFinderDirect}~\cite{Thornburg2004}, and perform Fourier analysis of timeseries as described in~\cite{manikantan_effects_2025}.
    
    We also report the dimensionless magnetic flux ($\phi_{\rm BH}$) threading each horizon. In coordinates where the horizon is a coordinate sphere and known apriori, such as the Kerr metric~\cite{tchekhovskoy_mad_2011}, or superposed Kerrschild metrics~\cite[see, e.g.,][]{ressler_dual_2025}, the magnetic flux integral is given by
    \begin{equation} \label{eq:og-flux}
        \Phi_{\rm BH} = \frac{1}{2} \int_{\theta} \int_{\varphi} |\mathcal{B}^r| dA_{\theta \varphi},
    \end{equation}
    where $dA_{\theta \varphi} = \sqrt{-g} \, d\theta \, d\varphi$ and $\sqrt{-g}$ is the determinant of the 4-metric. Here $\mathcal{B}^r$ is a radial magnetic field variable which equals $^*F^{0r}$, where $^*F$, is the Hodge dual of the electromagnetic field strength 2-form or Faraday tensor $F$. This magnetic field variable is not one measured by either a normal observer or an observer comoving with the plasma. 
    
    In dynamical spacetimes, the magnetic flux calculation is a not as simple, because the horizons are not spherical and not known a priori. Instead, we must locate these BH apparent horizons. The definition of the magnetic flux over a 2-surface $S$ follows from Maxwell's equations by analogy to the definition of the electric flux which gives the charge enclosed by a surface\footnote{The definition for the magnetic flux stems from Maxwell's homogeneous equation $dF=0$, where $d$ is the exterior derivative}
    \begin{equation}\label{eq:Bfluxdef}
        \Phi_{\rm BH} = \int_S F.
    \end{equation}
    The decomposition of $F$ in terms of electric and magnetic fields measured by an observer with 4-velocity equal to the unit  vector normal to spacelike hypersurfaces $n^\mu$ is~\cite{paschalidis_matching_2013} given by
    \begin{equation}\label{Fmndecomp}
     F_{\mu\nu}=n_\mu E_\nu -n_\nu E_\mu - \epsilon_{\mu\nu\alpha\beta}B^\alpha n^\beta,
    \end{equation}
    where $E^\mu=n_\nu F^{\mu\nu}$, $B^\mu=-n_\nu {}^*F^{\mu\nu}$ the electric and magnetic fields measured by a normal observer\footnote{Notice that in spherical polar coordinates $B^r=\alpha\, {}^*F^{0r}=\alpha\mathcal{B}^r$ with $\alpha$ the lapse function.}, and $\epsilon_{\mu\nu\alpha\beta}$ is the 4D Levi-Civita tensor. Here we adopt the convention that $\epsilon_{0123}=\sqrt{-g}$. Plugging Eq.~\eqref{Fmndecomp} into Eq.~\eqref{eq:Bfluxdef} and performing the integral over a BH apparent horizon surface, $\mathcal{S}$, one obtains
    \begin{equation}
      \label{eq:isolated-flux}
      \Phi_{\rm BH} = \int_{\mathcal{S}} \frac{1}{2!} \epsilon_{abc} B^c \d x^a \wedge \d x^b \,,
    \end{equation}
    where $\epsilon_{abc}=n^{\mu}\epsilon_{\mu abc}$ is the 3D Levi-Civita tensor for which $\epsilon_{123}=\sqrt{\gamma}$, where $\gamma$ is the determinant of the 3-metric on spatial hypersurfaces. If one where to perform the above integral over the closed surface of the BH apparent horizon, the integral would be zero, as there are no magnetic monopoles.  
    For this reason, we compute the surface integral of the absolute value of the magnetic flux, and divide by two to account for double counting of the flux on the bottom and top halves of the horizon, as in the original definition~(Eq. \eqref{eq:og-flux})\footnote{Given that $\alpha\sqrt{\gamma}=\sqrt{-g}$ Eq.~\eqref{eq:isolated-flux} reduces to Eq.~\eqref{eq:og-flux} in spherical polar coordinates after taking the absolute value and dividing by 2.}.  We now follow~\cite{bozzola_initial_2019}, who perform a similar integral to compute the electric charge of a BH. For completion, we outline the steps of our calculation here. The magnetic flux integral is coordinate-independent.
    Therefore, we choose cartesian coordinates, as is convenient for our simulations, $(x^a) = (x,y,z)$, and write the integral: 

    \begin{equation} \label{eq:phiB_full}
      \Phi_{\rm BH} = \frac{1}{2} \int_{\mathcal{S}} \sqrt{\gamma} \left
        |B_z \d x \wedge \d y + B_x \d y \wedge \d z - B_y \d x \wedge \d z \right|\,.
    \end{equation}
    Now, we introduce a parametrization for the horizon surface $\mathcal{S}$ in spherical polar coordinates where the horizon center is $(x_0, y_0, z_0)$:

    \begin{equation}
      \label{eq:parametrization-s}
      \begin{cases}
        x(\theta, \varphi) = x_0 + r(\theta, \varphi) \sin\theta \cos\varphi \\
        y(\theta, \varphi) = y_0 + r(\theta, \varphi) \sin\theta \sin\varphi \\
        z(\theta, \varphi) = z_0 + r(\theta, \varphi) \cos\theta
      \end{cases}
      \,,
    \end{equation}
    where $r(\theta, \varphi)$ is a suitable smooth function. Notice that this is the coordinate transformation between cartesian and spherical polar coordinates except $r\equiv r(\theta, \varphi)$ since our horizon is not spherical. The exterior (or wedge) products in Equation~\eqref{eq:phiB_full} are given by:

    \begin{equation}
        \d x \wedge \d y = |\rm{det} \, J_{xy}(\theta, \varphi)| \d \theta \d \varphi. 
    \end{equation}

    where the generic expression for the Jacobian, $J_{xy}$ is given by:

    \[
      J_{xy}(\theta, \varphi) =
        \begin{pmatrix}
        \partial_\theta x(\theta, \varphi)      & \partial_\varphi x(\theta, \varphi)  \\
        \partial_\theta y(\theta, \varphi)      & \partial_\varphi y(\theta, \varphi)  \\
        \end{pmatrix}
      \,.
    \]
    
    The first term in the integral of Equation~\eqref{eq:phiB_full}, therefore, would be evaluated as such:

    \begin{multline}
      \label{eq:integral-F-xy}
       \sqrt{\gamma} B_z(x,y,z) \d x \wedge \d y \\=   \sqrt{\gamma}  B_z(\theta, \varphi) \lvert \det J_{xy}(\theta,\phi) \rvert \d \theta \d \varphi\,,
    \end{multline}
    and so on, for the other terms.

    To obtain the dimensionless (and BH mass- and density-scale invariant) magnetic flux on the apparent horizon $\phi_{\rm BH}$, we normalize $\Phi_{\rm BH}$ by the rest-mass accretion rate and the gravitational radius of the BH
    \begin{equation} \label{eq:dimensionless_flux}
        \phi_{\rm BH} = \Phi_{\rm BH} / (\langle \dot{M} \rangle r_g^2 c)^{1/2},
    \end{equation}
    where $\langle \dots\rangle$ indicates a time-average, $r_g = GM/c^2$, where in our case each BH has $M=1/2$ with $G=c=1$. 
    
    We implement these equations in a modified version of \texttt{QuasiLocalMeasures} (QLM) and \texttt{QuasiLocalMeasuresEM}~\cite{bozzola_initial_2019}, which we call \texttt{QuasiLocalMeasuresMHD}\footnote{\href{https://github.com/arizona-relativity/QuasiLocalMeasuresMHD}{https://github.com/arizona-relativity/QuasiLocalMeasuresMHD}}. Our thorn is an extension of QLM that introduces an additional dependency on \ilgrmhd. It is straightforward to modify it to work with other thorns that evolve the magnetic field. 

\newpage
\bibliography{main.bib}

\end{document}